\documentclass[prb,floatfix,showpacs,superscriptaddress,amsmath,amssymb]{revtex4}
\usepackage{graphicx}
\usepackage{dcolumn}
\usepackage{bm}
\usepackage{color}
\usepackage{pst-all}

\begin{document}

\title{The spin-Peierls transition beyond the adiabatic approximation}

\author{A.O.\ Dobry}

\affiliation{Facultad de Ciencias Exactas Ingenier\'{\i}a y Agrimensura, Universidad Nacional de Rosario
and Instituto de F\'{\i}sica Rosario, Avenida 27 de Febrero 210 bis, 2000 Rosario, Argentina.}

\author{D.C.\ Cabra}

\affiliation{Laboratoire de Physique Th\'{e}orique, Universit\'{e} Louis Pasteur, 3 Rue de
l'Universit\'{e}, 67084 Strasbourg, C\'edex, France.}

\affiliation{Facultad de Ingenier\'\i a, Universidad Nacional de Lomas de Zamora, Cno.\ de Cintura
y Juan XXIII, (1832) Lomas de Zamora, Argentina.}

\affiliation{Departamento de F\'{\i}sica, Universidad Nacional de La Plata, C.C.\ 67, 1900 La
Plata, Argentina.}

\author{G.L.\ Rossini}

\affiliation{Departamento de F\'{\i}sica, Universidad Nacional de La Plata, C.C.\ 67, 1900 La
Plata, Argentina.}

\date{\today}

\begin{abstract}

We develop a theory of the spin-Peierls transition taking into account the three dimensional
character of the phonon field. Our approach does not rely on the adiabatic or mean field treatment
for the phonons. It is instead based in the exact integration of the phonon field,
the exact long wavelength solution of the one chain spin problem, and then a mean
field approximation for the interchain interaction. We show that the spin gap and the critical
temperature are strongly reduced due to the finite frequency effects of the phonon coupling
transverse to the magnetic chains. We claim that our results should be applicable to the inorganic
spin-Peierls compound CuGeO$_3$. We show that the long standing discussion on absence of a soft
mode in this compound can be naturally resolved within our theory.

\end{abstract}


\maketitle

\section{Introduction}

Quasi-one-dimensional electronic and magnetic materials are fascinating objects which have been
intensively studied  during the last decades. They show a variety of different phases as the
magnetically dimerized, unconventional metal and triplet superconductor phases. It is possible to
go from one phase to another by changing the experimentally accessible macroscopic parameters:
pressure, temperature and the applied magnetic field. The different phases arise from the
competition between the Coulomb repulsion between electrons and their coupling with the lattice
degrees of freedom. Whereas the essential physics relies on the one-dimensionality of the electronic
dynamics, the stabilization of the different phases depends on the three dimensional character of
the system and thus on the interchain interaction. Transitions between the different phases are, in
a way, a dimensional crossover from 1D to the three dimensionality of any real material.

Spin-Peierls systems are probably the simplest example of quasi-one-dimensional materials. They
show a high temperature phase which can be well modelled by a one dimensional antiferromagnetic
Heisenberg chain and a low temperature phase with lattice dimerization and spin singlet ground
state. The transition between them arises from the competition between the gain in
antiferromagnetic energy and the loss in elastic energy. It is indeed possible due to the three
dimensional character of the phonon field. It is therefore very attractive to carefully understand
the precise character of this transition as a representative of a large class of
quasi-one-dimensional materials.

The theory of spin-Peierls (SP) transition was developed in the 1970's in connection with organic
quasi-one-dimensional materials. The seminal work of Cross and Fisher\cite{CF} (CF) is nowadays
taken as the canonical theory of the spin-Peierls transitions. It is based on a RPA approximation
for the phonon field, which corresponds to neglecting the dynamics of the phonons. Otherwise the low
energy spin correlations are exactly taken into account by the bosonization method. The CF theory
is expected to work when the phonon frequency is much smaller than the spin gap. Moreover, CF have
shown that corrections to RPA are small if the bare phonon frequency in the direction of the
magnetic chain is much smaller than the phonon frequency perpendicular to the chains, that is to
say if there is a {\em preexisting soft mode}  before coupling phonons with the magnetism.

In this context the discovery of the first inorganic SP compound CuGeO$_3$ opened the possibility
to experimentally study the dimerized phase and the phase transition with unprecedented precision
\cite{Hase}. The recent characterization of a new family of spin-Peierls materials with an
incommensurate intermediate phase has renewed the interest on a precise description of the phase
transition\cite{TiClO}. The picture that emerges from neutron scattering measurement of the phonon
spectra in CuGeO$_3$ is the following\cite{Braden1,Braden2}: at room temperature, there are two
phonon modes relevant for the SP transition with $\omega_1=3.12\, THz$ and $\omega_2=6.53\, THz$ at the
(0.5 0 0.5) point of the Brillouin zone. The phonon dispersion transverse to the propagation
vectors of these relevant phonons is rather flat\cite{Braden2}. Therefore, {\em no preexisting
soft} mode is observed and the RPA is not a priori justified. Moreover, on lowering the
temperature, $\omega_1$ and  $\omega_2$ do not soften at the SP transition, as expected in the CF
scenario. This is a strange behavior for a displacive structural phase transition.

On the other hand, Gros and Werner\cite{GW} (GW) have identified a parameter regime  where
the CF approach does not predict a softening of the relevant phonons
but instead the appearance of a central peak in the dynamical structure factor
at the transition point.
The temperature dependence of the Peierls active phonon
modes predicted by this analysis are in agreement with neutron
scattering determinations in CuGeO$_3$. However the applicability
of the adiabatic approximation or RPA in this material is not a priori
justified.

In fact, what is missing for a complete understanding
of the spin-Peierls transition in CuGeO$_3$, and possibly in other
inorganic quasi-one-dimensional materials, is a theory which does
not rely on the static approximation for the phonons.
A first step in this direction was presented in [\onlinecite{ETD}],
where the dynamics of the transverse phonons in the antiadiadatic limit
was taken into account and the relevance the three dimensional character
of the phonons was first brought to light.
We give
some more elements of such a theory in the present work. 

The plan of the paper is as follows.
In Section II, we start
from a one-dimensional magnetic model coupled to two dimensional
phonons and integrate out the phonon coordinates. Then we treat the
effective in-chain interaction exactly by bosonization,
and the inter-chain interaction by a mean field approximation.
In Section III exact results for the 1D sine-Gordon model\cite{LZ} are used to solve the
mean field equations. By comparing with the adiabatic approximation we
identify the relevant parameter to analyze the deviations from the
adiabatic limit, namely the ratio between the frequency of phonon
relevant for the SP transition ($\omega_{\parallel}$) and the
frequency of the phonons transverse to the magnetic chain
($\omega_{\perp}$). The adiabatic limit is obtained only for high
enough value of $\frac {\omega_{\perp}} {\omega_{\parallel}}$.
Reducing this parameter both the gap and the spin-Peierls
temperature are reduced. For $\omega_{\perp} \to 0$, {\em i.e.}\  the pure
one dimensional problem, the gap and the SP temperature vanish. We
discuss our results in connection with a recent study of the AF
chain coupled to one-dimensional phonons\cite{COG}.
In Section IV we study the dynamics of the SP relevant phonon mode in the high temperature phase,
using a RPA on the effective interchain coupling
generated by the spin-phonon interaction.
We then address the question of absence of softening close to the SP transition.
Finally, in Section V we present our conclusions and prospects for future related work.

\section{The effective interactions induced by phonons}

Let us start by considering a two dimensional system of spin
$S=1/2$ antiferromagnetic Heisenberg chains, with exchange
constant $J_{\parallel}$ along one of the axes of a non-deformed
square elastic lattice. We label by $j$ chains, and by $i$ the
sites on each chain. The spin-phonon Hamiltonian is given by
\cite{dobry}:
\begin{eqnarray}
\label{hamsp2d}  H&=& \sum_{i,j}\frac{(P_{i}^{j})^2}{2m} +
\frac{1}{2}K_\parallel \sum_{i,j} (u_{i+1}^{j}-u_{i}^{j})^2
+\frac{1}{2}K_\perp \sum_{i,j} \left( (u_{i}^{j+1}-u_{i}^{j})^2
\right)+ J_\parallel \sum_{i,j} \left(1+\alpha
(u_{i+1}^{j}-u_{i}^{j})\right)\, \vec{S}_{i}^{j} \cdot
\vec{S}_{i+1}^{j} ,
\label{model}
\end{eqnarray}
where $u^j_i$ are the relevant (scalar) coordinates for ion displacements with respect
to equilibrium positions, $P^j_i$ are their conjugate momenta, $K_{\parallel}$ and $K_{\perp}$ are the elastic
couplings along in-chain and inter-chain ions, respectively, and $\alpha$ measures the deformation effect
on the magnetic exchange constants.

In the absence of phonons, we can write the bosonized low energy (long distance along the chains)
Hamiltonian for the Heisenberg spin chains\cite{CabraPujol} as
\begin{equation}
H_{spin}^j = J_\parallel \sum_{i,j} \vec{S}_{i}^{j} \cdot
\vec{S}_{i+1}^{j} \sim \frac{v_F}{2} \int {\rm d}x
\left(
K_L \left( \partial_x \tilde{\phi}^j(x) \right)^2
+
\frac{1}{K_L} \left(\partial_x \phi^j(x)\right)^2
\right),
\label{hbos}
\end{equation}
where $x=i a$ ($a$ being the lattice spacing), $\tilde{\phi}^j$ are the fields dual to the scalar
fields $\phi^j$, defined in terms of their canonical momentum as $\partial_x \tilde{\phi}^j =
\Pi^j$, and $:~:$ stands for normal order with respect to the isolated chains ground state. The
Fermi velocity of left and right excitations and the usual Luttinger parameter $K_L$ depend on the
$XXZ$ anisoptropy parameter, taking the values $v_F=\frac{\pi}{2} J_\parallel a$ and $K_L=1$ for
the isotropic Heisenberg chain with only  nearest neighbor interaction. A marginal term
$\lambda :\cos(2\sqrt{2\pi}\phi(x)): $
has been
discarded in eq.(\ref{hbos}).

The term coupling spins with phonons can be treated perturbatively
by using the continuum expression for the spin energy density
\begin{equation}
\vec{S}_{i}^{j} \cdot \vec{S}_{i+1}^{j} \sim \rho^j(x) =\frac{1}{\sqrt{2\pi}}  \partial_x \phi^j(x)
 + (-1)^i \beta : \cos(\sqrt{2\pi}\phi^j(x)) : + \cdots
\label{robos}
\end{equation}
where $\beta$ is a non-universal constant and the
ellipses indicate higher harmonics\cite{Haldane}. Notice that $: \cos(\sqrt{2\pi}\phi^j) : $ is
a relevant operator of scaling dimension $D=1/2$.

We will also assume that each chain dimerizes in such a way that the distortions can be
approximately described as $u_i^j \approx (-1)^i u^j(x)$. Notice however that neighboring chains
are not correlated {\em a priori}. Then
\begin{equation}
u^j_{i+1}-u^j_i \approx  (-1)^{i+1} ( 2u^j(x)+ a \partial_xu^j(x) ).
\end{equation}
We will keep only the leading order in the gradient expansion. The interaction Hamiltonian, in the
continuum limit and up to leading order in the lattice spacing, then reads
\begin{eqnarray}
H_{int}&=&\alpha  J_\parallel \sum_{i,j} (u_{i+1}^{j}-u_{i}^{j})\,
\vec{S}_{i}^{j} \cdot \vec{S}_{i+1}^{j}= -g \sum_j \int dx\,
u^j(x)\ : \cos(\sqrt{2\pi}\phi^j(x)) :\, ,
\end{eqnarray}
where $g \sim \frac{2 \alpha J_{\parallel} \beta}{a} $.

Regarding the phonon Hamiltonian, we find it convenient to first construct the corresponding Lagrangian.
Up to leading order in the lattice spacing the phonon Lagrangian
can be written in the continuum limit as
\begin{eqnarray}
L_{ph}=\frac{m}{2 a} \sum_j \int dx \left[ (\partial_t u^j(x,t))^2 -
(\omega_{\parallel}^2+\frac{1}{2}\omega_{\perp}^2)(u^j(x,t))^2+
\frac{1}{2}\omega_{\perp}^2 u^j(x,t) u^{j+1}(x,t) \right],
\end{eqnarray}
where $\omega_{\parallel}^2 = 4 \frac{K_{\parallel}}{m}$ and $\omega_{\perp}^2 = 4 \frac{K_{\perp}}{m}$.

Now we can write the action for the complete system and perform the
usual Wick rotation $\tau = i t$, arriving to the Euclidean action
describing the low energy physics at finite temperature $T$:
\begin{eqnarray}
S_E &=& \sum_j S_{spin}[\phi^j] \nonumber \\
&+& \frac{m}{2a} \sum_j \int dx \ d\tau
\left((\partial_\tau
u^j(x,\tau))^2 +(\omega_\parallel^2 +
\frac{1}{2}\omega_\perp^2)(u^j(x,\tau))^2 -
\frac{1}{2}\omega_\perp^2 u^{j+1}(x,\tau)u^j(x,\tau)\right)
\nonumber \\
&-& g \sum_j \int dx \ d\tau  \, u^j(x,\tau)\
:\cos(\sqrt{2\pi}\phi^j(x,\tau)):, \label{Lbos}
\end{eqnarray}
where $j$ runs from $1$ to $N$, $x$ from $0$ to $L$, $\tau$ from $0$ to $\beta=1/T$, and the spin
Euclidean action reads
\begin{equation}
S_{spin}[\phi^j]=
\frac{1}{2} \int  dx \ d\tau \left(
\frac{1}{v_F}(
\partial_\tau \phi^j(x,\tau)
)^2 +  v_F(\partial_x \phi^j(x,\tau))^2\right).
\label{Sbos}
\end{equation}

After Fourier transforming $j \to q$ (the wave vector perpendicular
to the magnetic chains) , $\tau \to \omega_n$, with
$\omega_n=\frac{2\pi n}\beta$ the Matsubara frequency, the
phonon Green's function reads:
\begin{eqnarray}
G^{jj'}(\tau,\tau')&=&
\frac{a}{m} \frac{1}{\beta}
\sum_n \int_{-\pi}^\pi \frac{dq}{(2\pi)}
\frac{e^{i q(j-j')} e^{-i\omega_n(\tau-\tau')}} {\omega_n^2+\omega(q)^2} \nonumber\\
&=& \frac{a}{2m} \int_{-\pi}^\pi \frac{dq}{2\pi} \frac{e^{i
q(j-j')}}{ \omega(q)}\left[ e^{-\omega(q)|\tau-{\tau}'|} + 2
N(\omega(q)) \cosh(\omega(q)({\tau}'-\tau))\right],
\label{afterphonons}
\end{eqnarray}
where $N(\omega(q))$ is the Bose distribution and the phonon
dispersion relation is given by $\omega^2(q)=\omega_{\parallel}^2+
\omega_{\perp}^2 \sin(q/2)^2$.
The action is quadratic in the phonon field and can be easily
integrated, rendering the effective action for coupled spin
chains
\begin{equation}
S_{eff} = \sum_j S_{spin}[\phi^j] -\frac{g^2}{2} \sum_{j,j'} \int dx
\int d\tau \ d\tau' \left(:\cos(\sqrt{2\pi}\phi^j(x,\tau)):
G^{jj'}(\tau,\tau') :\cos(\sqrt{2\pi}\phi^{j'}(x,\tau')): \right).
\label{Lbos2a}
\end{equation}
We find it convenient to separate the in-chain from the inter-chain interactions writing
\begin{equation}
S_{eff} = \sum_j S_{spin}[\phi^j] + S_{in} + S_{inter}
\label{Seff}
\end{equation}
where
\begin{equation}
S_{in} = -\frac{g^2}{2} \sum_{j} \int dx \int d\tau \ d\tau'
\left(:\cos(\sqrt{2\pi}\phi^j(x,\tau)):  G^{jj}(\tau,\tau') :
\cos(\sqrt{2\pi}\phi^{j}(x,\tau')): \right)
\label{Sin}
\end{equation}
and
\begin{equation}
S_{inter} = -\frac{g^2}{2} \sum_{j\neq j'} \int dx \int d\tau \
d\tau' \left(:\cos(\sqrt{2\pi}\phi^j(x,\tau)): G^{jj'}(\tau,\tau')
:\cos(\sqrt{2\pi}\phi^{j'}(x,\tau')): \right).
\label{Sinter}
\end{equation}

We will now treat the inter-chain interaction with a mean field
decoupling.
Following Delfino {\em et.\ {al.}}\cite{ETD}, we write
\begin{equation}
:\cos(\sqrt{2\pi}\phi^j): =
\epsilon_0 + N(\cos(\sqrt{2\pi}\phi^j)),
\label{hocuspocus}
\end{equation}
where the normal order $N$ is now taken with respect to the vacuum of the coupled chains system.
Here $\epsilon_0$ represents the subtraction of the staggered energy density
($\epsilon_0(x=ia)\sim (-1)^{i+1}
(<\vec{S}_i^j\cdot\vec{S}_{i+1}^j>- <\vec{S}_{i+1}^j\cdot\vec{S}_{i+2}^j>)$\,) in
the coupled system, describing the magnetic dimerization along chains. According to the
dimerization picture, we take $\epsilon_0$ to be position independent. Then
\begin{eqnarray}
S_{inter} &=&
-\frac{g^2}{2}
\sum_{j \neq j'} \int dx \int d\tau \ d\tau'
\left\{ \epsilon_0^2 \ G^{jj'}(\tau,\tau') +
\epsilon_0 N(\cos(\sqrt{2\pi}\phi^{j}(x,\tau)))\ G^{jj'}(\tau,\tau') \right.
\nonumber\\
&+&
\left.\epsilon_0 N(\cos(\sqrt{2\pi}\phi^{j'}(x,\tau')))\ G^{jj'}(\tau,\tau') + \cdots
\right\} \label{Sinter2}
\end{eqnarray}
where the ellipsis corresponds to quadratic fluctuations in the
inter-chain coupling, which we will neglect at the MF level
({\em i.e.}\ we assume that fluctuations are small with respect to
the expectation value $\epsilon_0$).
Due to isotropy and translation symmetry
$G^{jj'}(\tau,\tau')=G^{|j-j'|}(|\tau-\tau'|)$, then the last two
terms are equal. Moreover, we can change the summation indices
from $j , j'$ to $j , J=j'-j$ and $\tau, \tau'$ to $\tau,
\Theta=\tau'-\tau$ and perform the sum over $J$ and the integral over
$\Theta$. We obtain
\begin{eqnarray}
\int_0^\beta d\tau' \sum_{J\neq 0} G^{J}(|\tau-\tau'|)
&=&
\int_0^\beta d\Theta \left(\sum_{J} G^{J}(\Theta) - G^{0}(\Theta) \right) \nonumber\\
&=&
\frac{a}{2m}\{\frac{1}{\omega_{\parallel}^2}\left[(1-e^{-\beta\omega_{\parallel}})+2
N(\omega_{\parallel}) \sinh(\beta \omega_{\parallel})\right] \nonumber\\
& & -
\frac{1}{2}  \int_{-\pi}^\pi \frac{dq}{2\pi} \ \frac{1}{\omega(q)^2}
\left[(1-e^{-\beta\omega(q)}) + 2 N(\omega(q)) \sinh(\beta \omega(q))\right]\},
\label{interGreen}
\end{eqnarray}
notably leading to the temperature independent result
\begin{eqnarray}
\int_0^\beta d\tau' \sum_{J\neq 0} G^{J}(|\tau-\tau'|) &=&
\frac{a}{m}\frac{1}{{\omega_\parallel}^2} \left( 1-\frac{1}{
\sqrt{1+\frac{\omega_\perp^2}{\omega_\parallel^2}}  } \right)
\equiv F(\omega_{\parallel},\omega_{\perp}).
\label{factorAriel}
\end{eqnarray}
Then, $S_{inter}$ reads
\begin{eqnarray}
S_{inter}
&=&
-\epsilon_0^2 \, \frac{g^2}{2}
\sum_{j}  \int dx \ d\tau \ F(\omega_{\parallel},\omega_{\perp}) \nonumber \\
&& -\epsilon_0 \, g^2 \sum_{j} \int dx \int d\tau \
N(\cos(\sqrt{2\pi}\phi^{j}))\ F(\omega_{\parallel},\omega_{\perp}).
\label{Sinter3}
\end{eqnarray}
After writing the space-time volume integral, and writing back in
the last term $N(\cos(\sqrt{2\pi}\phi^{j}(\tau))=:\cos(\sqrt{2\pi}\phi^{j}(\tau)):
-\epsilon_0$,
the effective action in eq.(\ref{Seff}) takes the mean field
form
\begin{eqnarray}
S_{eff}^{\cal MF}
&=&
\sum_j S_{spin}[\phi^j]
-
\frac{g^2}{2} \sum_j \int dx \ d\tau \ d\tau'
:\cos(\sqrt{2\pi}\phi^{j}(x,\tau)): G^{0}(\tau,\tau') :\cos(\sqrt{2\pi}\phi^{j}(x,\tau')):
\nonumber \\
&  &
- \epsilon_0 \, g^2
F(\omega_{\parallel},\omega_{\perp}) \sum_{j} \int dx \ d\tau \ :\cos(\sqrt{2\pi}\phi^{j}(x,\tau)):
+\frac{1}{2} \epsilon_0^2 \, g^2
F(\omega_{\parallel},\omega_{\perp}) N L \beta .
\label{SeffMF}
\end{eqnarray}
The second term in the first line is the remaining in-chain retarded
interaction. In the antiadabatic limit ($m \rightarrow \infty$) it
becomes an instantaneous interaction which is a marginally
irrelevant perturbation.
Following an argument similar as the one presented in [\onlinecite{FH}]
(see the discussion starting from eq.(2.30)), it is possible to
show that even in the finite mass case this term gives rise to an
irrelevant operator (see also [\onlinecite{COG}]). Consistently with the bosonized spin Hamiltonian in eq.(\ref{hbos}),
we neglect it in the following.

Finally we obtain
\begin{eqnarray}
S_{eff}^{\cal MF}
&=&
\sum_j S_{spin}[\phi^j]
-
\epsilon_0 \, g^2
F(\omega_{\parallel},\omega_{\perp}) \sum_{j} \int dx \ d\tau \ :\cos(\sqrt{2\pi}\phi^{j}(x,\tau)):
+
\epsilon_0^2 \, \frac{g^2}{2}
F(\omega_{\parallel},\omega_{\perp}) N L \beta .
\label{Snonad}
\end{eqnarray}
The second term is a strongly relevant perturbation that opens a gap
for any value of the phonon frequencies $\omega_{\parallel}$ and
$\omega_{\perp}$, as long as $\epsilon_0$ does not vanish
(this will be evaluated by a self-consistent procedure in the next section).
Note that
this  behavior is different than the one obtained
in a pure one-dimensional model\cite{COG}. In that case, going
from the adiabatic to the antiadiabatic limit, the gap closes at an
intermediate value of the phonon frequency. Our previous discussion
shows that when the 2 or 3D character of the phonons is taken
into account such a crossover does not show up. In [\onlinecite{ETD}]
a model similar to (\ref{model}) has been considered only in the
antiadiabatic limit; our results show that their fully gapped phase
extends over a finite frequency range. In addition, as we have
kept the explicit dependence on the phonon frequencies, we can
analyze the evolution of the gap and the SP transition temperature.
We will undertake this analysis in the following section.

\section{The spin gap and transition temperature as a function of the phonon frequency}

Let us start by discussing how the Mean Field adiabatic treatment of the
phonon field can be implemented in our calculation. This will
give results for the spin gap and the transition temperature
generalizing those in [\onlinecite{OC}], which are in fact extensions of the classical
work of Cross and Fisher\cite{CF}.
This calculation will  set up the energy and
temperature scale to be compared with our more general
non-adiabatic results. Fixing the displacement field in eq.(\ref{Lbos}) to be a constant
(chain, position and time independent) $u^j(x,\tau)=u$, we have
\begin{eqnarray}
 S_{ad} = N \left[ S_{spin}[\phi] - g u \int dx d\tau
:\cos(\sqrt{2\pi}\phi(x,\tau)): +  \frac{m}{2 a} L \beta\omega_\parallel^2 u^2\right].
\label{Sad}
\end{eqnarray}
Note that the term with $\omega_{\perp}$ vanishes and therefore the total action is
$N$ times the action of each chain.

We obtain $u$ in a self-consistent way by minimizing the total
energy with respect to it. The ground state energy is evaluated
using the exact results for the massive sine-Gordon theory.  The lowest energy excitation is a
soliton of mass $M_{ad}$, which can be obtained from eq. (12)
of [\onlinecite{LZ}] ( $2 \mu=gu/v_F$ and $\beta=1/2$ in our notation):
\begin{eqnarray}
\frac{gu}{2 v_F}= \frac{\Gamma(\frac14)}{\pi \Gamma(\frac34)}
\left[ M_{ad} v_F\sqrt{\frac{\pi}{4}}\frac{\Gamma(\frac23)}
{\Gamma(\frac16)} \right]^\frac32 .
\label{relgmu}
\end{eqnarray}
The specific ground state energy for the sine-Gordon theory is  given by
$-\frac14 M_{ad}^2 v_F^3 \tan (\frac{\pi}{6})$, so that the total energy density reads
\begin{eqnarray}
e^{ad} \equiv \frac{E^{ad}}{NL}=
\frac{m}{2a} \omega_{\parallel}^2 u^2-\frac14 M_{ad}^2 v_F^3 \tan (\frac{\pi}{6}).
\label{Ead}
\end{eqnarray}
Inverting (\ref{relgmu}) to obtain $M_{ad}(u)$ and minimizing (\ref{Ead}) with respect to $u$ we get:
\begin{eqnarray}
M_{ad}=\frac{g^2}{4 v_F^2} \frac{a}{m \omega_\parallel^2} \frac13 \tan(\frac{\pi}{6})
\left(\frac{\pi \Gamma(\frac34)}{\Gamma(\frac14)} \right)^2 \left(\sqrt{\frac{4}{\pi}}
\frac{\Gamma(\frac16)}{\Gamma(\frac23)}\right)^3
\equiv H \frac{g^2}{v_F^2}\frac{a }{m \omega_\parallel^2},
\label{Mad}
\end{eqnarray}
where $H=5.4133$ is just a constant collecting all of the numerical factors.
The spin gap $\Delta^{ad}$  is then given by
\begin{eqnarray}
\Delta^{ad}= M_{ad} v_F^2.
\label{deltaad}
\end{eqnarray}
Equations (\ref{Mad}) and (\ref{deltaad}) give the zero temperature gap in terms of the model
parameters $g$, $J_\parallel$ and $\omega_{\parallel}$. It is equivalent to the first of the eqs.\ (8) in
\cite{OC}.

In the Mean Field adiabatic treatment, the spin-Peierls transition temperature
$T_{SP}^{ad}$ can be obtained by considering
the finite temperature free energy. Borrowing the appropriate expansion in powers of $u$ from
Eq. (12) of [\onlinecite{OC}] ($\frac{2g\delta}{(2\pi a)^2}$ must be replaced by
$gu/\sqrt{a}$) the lowest order of the sine-Gordon model free energy density reads
\begin{eqnarray}
-\frac{\pi}{6 v_F}T^2 - \frac{\pi^2}{4\Gamma^4(\frac34)}\frac{(gu)^2 }{T}  + O(u^4),
\label{fmgad}
\end{eqnarray}
so that the full variational free energy is obtained by adding the contribution of the elastic energy,
\begin{eqnarray}
f^{ad} \equiv \frac{F^{ad}}{NL}=
\frac{m}{2 a} \omega^2_{\parallel} u^2-\frac{\pi}{6 v_F}T^2 -
\frac{\pi^2}{4\Gamma^4(\frac34)}\frac{(gu)^2 }{T}  + O(u^4).
\label{fad}
\end{eqnarray}
The transition temperature corresponds to the one where the free energy minimum shifts from
$u \neq 0$ to $u=0$.
Although we have not written quartic terms in $u$, it is easy to see that the transition
is signaled by a change of sign in the coefficient of $u^2$.
This condition gives
\begin{eqnarray}
T^{ad}_{SP}=\frac{a g^2 \pi^2}{2  \Gamma^4(\frac34) m
\omega^2_{\parallel}}.
\end{eqnarray}

We now undertake the calculation of the gap and the transition temperature in the  non adiabatic case.
We start from the action (\ref{Snonad}), our main result previously obtained.
Comparing eqs.\ (\ref{Snonad}) and (\ref{Sad}) we note that
the soliton mass $M$ can be computed by following similar steps as we did to obtain $M^{ad}$.
Now, the mean field variational parameter is $\epsilon_0$ instead of $u$ (indeed
the factor  $g F \epsilon_0$ plays the role of $u$); the total energy density is given by
\begin{eqnarray}
e \equiv \frac{E}{NL}=\frac{\epsilon_0^2 g^2}{2}F(\omega_{\parallel},\omega_{\perp}) - \frac14 M^2 v_F^3 \tan(\frac{\pi}{6}).
\end{eqnarray}
We readily obtain the soliton mass as
\begin{eqnarray}
M = H \frac{g^2}{v_F^2} F(\omega_{\parallel},\omega_{\perp}),
\label{M}
\end{eqnarray}
and the corresponding gap $\Delta = M v_F^2$.
Comparing this result with eq.(\ref{Mad}) we obtain the relation between the
gap in the non-adiabatic  and the adiabatic case:
\begin{eqnarray}
\frac{\Delta}{\Delta_{ad}}=1-\frac{1}{\sqrt{1+(\frac{\omega_\perp}{\omega_\parallel})^2}}.
\label{relMMad}
\end{eqnarray}

\begin{figure}[htp]
\vspace{2mm}
\includegraphics[width=0.40\textwidth]{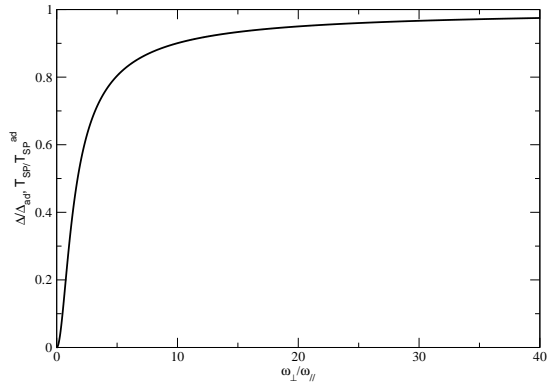}
\caption{\label{relgap} Dependence of the gap and the SP
temperature with $\frac{\omega_{\perp}}{\omega_{\parallel}}$ .}
\end{figure}

In Fig. \ref{relgap} we show the evolution of the gap with
the ratio $\frac{\omega_\perp}{\omega_\parallel}$. Only for
$\frac{\omega_\perp}{\omega_\parallel}\rightarrow \infty$, {\em i.e.}\
when the lines perpendicular to the chain move rigidly, the gap
corresponds to the adiabatic one. Note that this is precisely the
condition established by CF for the validity of their RPA
calculations. In the general case the gap decreases when
$\frac{\omega_\perp}{\omega_\parallel}$ decreases. This is an
important result applicable to a general non-adiabatic spin-Peierls
system. For example, for CuGeO$_3$ we can take as the relevant
phonon mode (with frequency $\omega_{\parallel}$) the projections
of the (0.5 0 0.5) phonons on the direction of the magnetic
chains. The transverse direction is then the
$b$-direction along the path (0.5 x 0.5). The phonon dispersion
relation $\omega^2(q)=\omega_{\parallel}^2+ \omega_{\perp}^2
\sin(q/2)^2$ in our model corresponds to phonons in this direction. We
estimate $\omega_{\perp}$ from the frequencies at the center and
at the boundary of the Brillouin zone in this direction. It has
been recognized that there are two $T^2_+$ phonons at (0.5 0 0.5)
which are the relevant for the SP distortion (with frequencies
$\omega_1=3.12\, THz$ and $\omega_2=6.53\, THz$ at room temperature)\cite{Braden1}. From
Fig.\ 12 in [\onlinecite{Braden2}] we can follow the dispersion of the
branches in the direction (0.5 x 0.5) connected with these two
modes.  The lowest energy branch is rather flat and we do not take
it into account. For the phonon with the frequency $\omega_2$ we
have $\frac{\omega_{\perp}}{\omega_{\parallel}}  \sim 0.58$ and
the gap from eq.\ (\ref{relMMad}) is reduced to $13\%$ of the one estimated by
the adiabatic approximation.
In practice, the measured zero temperature gap is used to
estimate the spin-phonon coupling\cite{rieradobry}. Our previous
result implies that the spin-phonon coupling could be strongly
underestimated by using adiabatic results.
Indeed, from our present approach we estimate the
dimensionless spin-phonon coupling ($\lambda=\frac{J\alpha^2}{K}$)
for CuGeO$_3$ one order of magnitude larger than the previously
calculated one. This could have important consequences to interpret
experimental results in this and other materials.

A relation similar to eq.\ (\ref{relMMad}) holds between the transition temperature in the
non-adiabatic and the adiabatic case.
The free energy
corresponding to action (\ref{Snonad}) should be expanded in powers
of $\epsilon_0$.  We take again this expansion  from Eq.\ (12) of
[\onlinecite{OC}] (now $\frac{2g\delta}{(2\pi a)^2}$ is replaced by
$\epsilon_0 g^2 F(\omega_\parallel,\omega_\perp)/\sqrt{a}$). The total free energy density is
\begin{eqnarray}
f \equiv \frac{F}{NL}
=
-\frac{\pi}{6 v_F}T^2+
\epsilon_0^2\frac{g^2 F(\omega_\parallel,\omega_\perp)}{2}
\left(1-\frac{g^2 \pi^2 F(\omega_\parallel,\omega_\perp)}{2 \Gamma^4(\frac34) T} \right)\, .
\end{eqnarray}
The vanishing of the factor proportional to $\epsilon_0^2$  signals the transition temperature. We get
\begin{eqnarray}
T_{SP}=\frac{g^2  \pi^2}{ 2\Gamma^4(\frac34)}F(\omega_\parallel,\omega_\perp)
\label{TSPMF}
\end{eqnarray}
and the promised relation is:
\begin{eqnarray}
\frac{T_{SP}}{T^{ad}_{SP}}=1-\frac{1}{\sqrt{1+(\frac{\omega_\perp}{\omega_\parallel})^2}}.
\label{relTsp}
\end{eqnarray}

Notice that the ratio $\frac {\Delta}{T_{SP}}$ from our computation does not depend on
$\frac{\omega_\perp}{\omega_\parallel}$. The value of such a
ratio seems to be fixed by the adiabatic calculation.
The existence of such universal ratio is similar to the BCS mean field
theory of superconductivity. Moreover, the experimental value
observed  in CuGeO$_3$  is close to the BCS-ratio $1.76$.  In fact
a BCS-like theory could be obtained from the original Heisenberg
model coupled to phonons by a Jordan-Wigner transformation and
neglecting interaction between the spinless fermions.
Note however
that the spinless fermions are strongly interacting, and this fact
renormalizes the BCS ratio away from the non interacting
value\cite{OC}. The precise value of this ratio is difficult to
obtain because it is affected by logarithmic corrections induced by
marginally irrelevant terms. Indeed, the effect of the marginal
term arising from the Heisenberg interaction is expected to be
small in CuGeO$_3$ because the second next-nearest neighbor is
closer to the critical value\cite{rieradobry}, where logarithmic
corrections vanish. However, as can be seem from action
(\ref{Seff}), logarithmic corrections are also expected from the
spin-phonon interaction term and they do depend on the phonon
frequencies. Therefore we expect that the exact value of $\frac
{\Delta}{T_{SP}}$ will ultimately depend on
$\frac{\omega_\perp}{\omega_\parallel}$.

\section{Dynamics of the phonon modes}
%

The original work of CF on spin-Peierls systems predicts the softening of a
phonon mode above the transition temperature, whose polarization
pattern corresponds to the static distortion leading to the low
temperature phase below the transition point $T_{SP}$. The softening is
due to the coupling of the phonons with the one-dimensional
magnetism. This is in fact a quite general scenario for a
structural displacive phase transition. Therefore it came as a
surprise that the SP transition in CuGeO$_3$ shows no phonon
softening\cite{Hirota,Lorenzo}.
Gros and Werner\cite{GW} have reanalyzed the
CF approach showing that, for higher enough bare phonon frequency,
there is a second pole of the phonon propagator without softening
of the active SP phonon, characterized by the presence of a central peak
in the dynamical structure factor near the
transition point. Moreover, the RPA approach underlying the CF
calculation is justified if the SP phonon has a bare frequency
considerably lower than other momentum phonons, {\em i.e.}\ a non-magnetic
softening observable at high temperature relative to the
SP transition temperature. This is not the situation in CuGeO$_3$ and
application of RPA seems questionable.

In the following we show how by extending our previous
calculations to the dynamics of the phonon modes we can solve this
puzzle. The extension of the  static MF approximation on the
interchain interaction to the dynamical correlation function is
accomplished by a RPA analysis along the lines of
[\onlinecite{Schulz,ETD}]. In contrast to the one-dimensional analysis in CF,
this RPA is
based on the {\em quasi}-one-dimensionality underlying this kind of
materials, and does not require a preexisting mode softening.
It becomes exact in the limit of an infinite number of
neighboring chains.

The dressed phonon frequencies are obtained from the poles of the
retarded phonon Green's function $D(k,q,\omega)$\cite{Mahan} which
can be computed from the Matsubara Green's function
$D(k,q,\omega_n)$, with wave vectors $k$ in the chain direction and $q$ the transverse one,
by analytic continuation as
$D(k,q,\omega)=D(k,q,i\omega_n \rightarrow \omega)$.
This is in
fact the Fourier transform of the Green's function in real space and
imaginary time $D^{j-j'}(x-x',\tau-\tau')$, which can be found by
functional derivation of the partition function when a current
term is added:
\begin{eqnarray}
Z[J]=\int {\cal D} u^j {\cal D} \phi^j \exp \left\{
- \sum_j\int dx\, d\tau [{\cal L}(\phi^j(x,\tau),u^j(x,\tau))+ g J^j(x,\tau)u^j(x,\tau)]\right\}.
\label{Z}
\end{eqnarray}
Integrating out the phonon coordinates as before we
obtain the following effective action
\begin{eqnarray}
S_{eff}&=&\sum_j S_{spin}[\phi^j]\nonumber\\
 &-& \frac{g^2}{2} \sum_{j,j'}
\int dx \int d\tau d\tau'
\left(:\cos(\sqrt{2\pi}\phi^j(x,\tau)):+J^j(x,\tau)\right)
G^{jj'}(\tau,\tau')
\left(:\cos(\sqrt{2\pi}\phi^{j'}(x,\tau')):+J^{j'}(x,\tau') \right).
\label{SeffJ}
\end{eqnarray}
The dressed phonon propagator
\begin{eqnarray}
D^{j-j'}(x-x',\tau-\tau')=\left.\frac{1}{g^2} \frac{\delta^2 \ln Z[J]}{ \delta
J^{j'}(x',\tau') \delta J^{j}(x,\tau)} \right|_{\{J^j(x,\tau)\}=0}
\label{D}
\end{eqnarray}
can be obtained from eqs.\ (\ref{Z}, \ref{SeffJ}, \ref{D}) as
\begin{eqnarray}
D^{j-j'}(x-x',\tau-\tau') &=&
G^{j'-j}(x'-x,\tau'-\tau)
-
g^2 \sum_{\hat{j},\hat{j'}}
\int d\hat{x}d\hat{x'}d\hat{\tau}d\hat{\tau'}
G^{\hat{j}-j}(\hat{x}-x,\hat{\tau}-\tau)
G^{\hat{j}'-j'}(\hat{x'}-x',\hat{\tau'}-\tau')\nonumber\\
&&\langle {\bf\cal T} \left(:\cos(\sqrt{2\pi}\phi^{\hat{j}}(\hat{x},\hat{\tau})):
:\cos(\sqrt{2\pi}\phi^{\hat{j'}}(\hat{x'},\hat{\tau'})):\right) \rangle ,
\end{eqnarray}
where $G^{j-j'}(x-x',\tau-\tau')= \delta(x-x') G^{jj'}(\tau,\tau')$ is
a convenient notation for the non-interacting phonon propagator and ${\bf\cal T}$ indicates Euclidean time-ordering.
After Fourier transforming to $(k,q,\omega_n)$ coordinates and performing the analytic
continuation $i\omega_n \rightarrow \omega$ we obtain the desired
retarded phonon Green's function
\begin{eqnarray}
D(k,q,\omega)=G(q,\omega) + g^2 G^2(q,\omega) \chi(k,q,\omega),
\label{retph}
\end{eqnarray}
where $G(q,\omega)=\frac{a}{m}\frac{1}{-\omega^2+\omega(q)}$ is the bare
phonon Green's function and $\chi(k,q,\omega)$ stems for the Fourier
transform of the retarded correlator of
$:\cos(\sqrt{2\pi}\phi^{j}):$ .
In order to evaluate bulk corrections arising from inter-chain coupling, we
will calculate this correlator using a RPA approach in the interchain interaction\cite{ETD}. It
reads
\begin{eqnarray}
\chi^{RPA}(k,q,\omega)=\frac{\chi^0(k,\omega;T)}{1-g^2G_{inter}(q,\omega)\chi^0(k,\omega;T)},
\label{chirpa}
\end{eqnarray}
where $\chi^0(k,\omega;T)$ is the already known finite temperature one chain $:\cos(\sqrt{2\pi}\phi):$-correlator
and $G_{inter}$ is the inter-chain bare phonon propagator,
defined in coordinate space by the subtraction
\begin{equation}
G_{inter}^{j-j'}(x-x',\tau-\tau')=
G^{j-j'}(x-x',\tau-\tau') -
\delta^{jj'} G^{0}(x-x',\tau-\tau').
\label{defGinter}
\end{equation}
The one chain $:\cos(\sqrt{2\pi}\phi):$-correlator at finite temperature has been computed
elsewhere\cite{CF,GW,Giamarchi}, being given by
\begin{eqnarray}
\chi^0(k,\omega;T) = \frac{2 d}{T} I_1(\frac{(\omega-\Delta)}{2\pi T})
I_1(\frac{(\omega+\Delta)}{2\pi T}),
\end{eqnarray}
where $d \sim 0.37$ is a constant weakly depending of the momentum
cutoff in the bosonization procedure, $\Delta=v_F  |k-\pi|$ is the
lower edge of the two spinon continuum, and
\begin{eqnarray}
I_1(k)=\frac{1}{2\pi}\int_0^{\infty}dx
e^{ikx}(\sinh(x))^{-\frac12}=\nonumber\\
\frac{1}{2\sqrt{2\pi}}\frac{\Gamma[\frac14-i\frac{k}{2}]}{\Gamma[\frac34-i\frac{k}{2}]}.
\end{eqnarray}
It will be useful to note that at low frequency one can expand
\begin{equation}
T\chi^0(k,\omega;T)\approx \chi_0 -i\chi_1 \frac{\omega}{2\pi T} +\chi_2 \left(\frac{\omega}{2\pi T}\right)^2,
\label{chi-expansion}
\end{equation}
with $\chi_0 \sim 0.26$ and $\chi_2 \sim 2.2$.

The Fourier transform of the
interchain propagator $G^{inter}$ is obtained (see Section II) as
\begin{eqnarray}
G^{inter}(q,\omega)&=&\frac{1}{-\omega^2+\omega(q)}-
\Upsilon(\omega),
\label{Ginter}
\end{eqnarray}
with
\begin{eqnarray}
\Upsilon(\omega)=\frac{1}{2\pi}\int
\frac{dq'}{\omega^2(q')-\omega^2}
=\frac{1}{\sqrt{(\omega_{\parallel}^2-\omega^2)}\sqrt{(\omega_{\parallel}^2+\omega_{\perp}^2-\omega^2}}.
\end{eqnarray}

We are interested in the temperature evolution of the SP active
phonon mode, given by $k=\pi$ and $q=0$. From
eq.(\ref{retph}-\ref{Ginter}) we have
\begin{eqnarray}
D(0,\pi,\omega)=\frac{g^2(1-g^2
\Upsilon(\omega)\chi^0(\pi,\omega;T))}{
G^{-1}+\frac{g^2}{2}(1-G^{-1}\Upsilon(\omega))\chi^0(\pi,\omega;T)}
\end{eqnarray}
and the phonon frequencies are given by the poles over the real
axis of this expression, {\em i.e.}\ the roots of the following
equation
\begin{eqnarray}
-\omega^2+\omega_{\parallel}^2+\frac{g^2}{2}(1-(\omega_{\parallel}^2-\omega^2)\Upsilon(\omega))
Re\chi^0(\pi,\omega;T)=0 .
\label{denzeros}
\end{eqnarray}

Let us start the analysis of the consequences of the previous
results by recalculating the spin-Peierls transition
temperature. Note that, different than the previous Section, we are
now coming from the high temperature phase to the transition
point. As discussed by GW, $T_{SP}$ is signaled by a macroscopic
occupation of the Peierls active phonon mode, {\em i.e.}\ the transition
temperature takes place when eq.\ (\ref{denzeros}) has a solution
$\omega=0$. We then obtain
\begin{eqnarray}
T_{SP}=\frac{\chi_0
g^2}{2 \omega_{\parallel}^2}\left(1-\frac{1}{\sqrt{1+\frac{\omega_{\perp}^2}{\omega_{\parallel}^2}}}\right).
\label{TSPRPA}
\end{eqnarray}
It is important to stress that eq.\ (\ref{TSPRPA}) gives the
same dependency of $T_{SP}$ on the microscopic parameters than
eq.\ (\ref{TSPMF}) in the previous Section; differences in the numerical prefactor
arise from different criteria to fix the momentum cutoff. Note
however that relation (\ref{relTsp}) is fulfilled. This results
prove that our RPA procedure contains the previous MF approach
when the static properties are considered.

\begin{figure}[htp]
\vspace{2mm}
\includegraphics[width=0.50\textwidth]{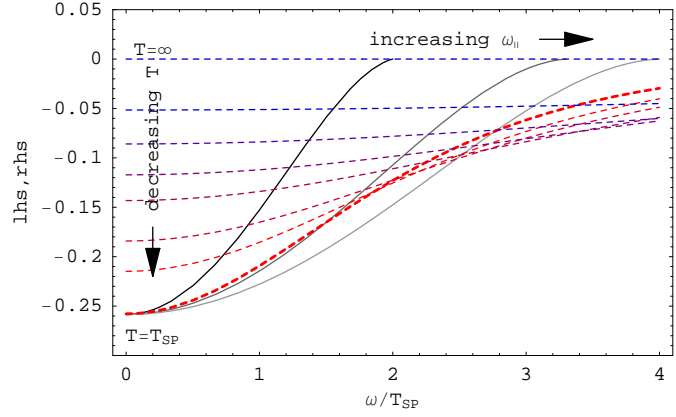}
\caption{
Evolution of the SP relevant phonon mode. Dashed lines, from blue to red,
represent the rhs of eq.\ (\ref{freres}) when
temperature is lowered until $T_{SP }$.
Solid gray lines represent the lhs for $\omega_\parallel/T_{SP} = 2,3,4$ and $\omega_\perp/T_{SP} = 1$.
The intersections illustrate the evolution of the renormalized phonon frequency.
For $\omega_\parallel/T_{SP} = 2$ the phonon softens completely,
for $\omega_\parallel/T_{SP} = 3$ it softens partially,
and for $\omega_\parallel/T_{SP} = 4$ it hardens close to $T_{SP}$.
}
\label{GW1}
\end{figure}

In the rest of this Section we face the question of phonon softening,
discussing the calculation of the renormalized phonon
frequencies. We use (\ref{TSPRPA}) to eliminate the cut-off dependent coupling $g^2$ from eq.\ (\ref{denzeros}).
After straightforward algebra we obtain
\begin{eqnarray}
\chi_0
\left(1-\frac{1}{\sqrt{1+(\frac{\omega_{\perp}} {\omega_{\parallel}})^2}}\right)
\left(\left(\frac{\omega}{\omega_{\parallel}}\right)^2-1\right) /
\left(1-\frac{\sqrt{1-(\frac{\omega}{\omega_{\parallel}})^2}}
{\sqrt{1+(\frac{\omega_{\perp}}{\omega_{\parallel}})^2+(\frac{\omega}{\omega_{\parallel}})^2}}
\right)
 =
 T_{SP} Re \chi^0(\pi,\omega). \label{freres}
\end{eqnarray}
This equation is the generalization of the GW results to finite values of
$\frac{\omega_{\perp}}{\omega_{\parallel}}$, and agrees with their corresponding eq.\ (6) when
$\frac{\omega_{\perp}}{\omega_{\parallel}}\gg 1$.
There are two different temperature dependence
regimes of the renormalized phonon frequency $\omega(\pi,0)$, as depicted in Fig.\ \ref{GW1}. For
low enough $\omega_{\parallel}$ the phonon progressively softens from higher $T$ until $T_{SP}$,
where it drops to zero frequency.
In this regime the phonon softening explains the SP transition.
For higher $\omega_{\parallel}$, the phonon frequency remains finite up to the SP temperature and
no soft phonon behavior is obtained; in fact, there can be partial softening or even hardening of
the renormalized frequency close to $T_{SP}$. However, an additional solution $\omega =0$
in eq.\ (\ref{freres}) signals the
appearance of a central peak in the dynamical structure factor, leading to the SP transition
\cite{GW}.

\begin{figure}[htp]
\vspace{2mm}
\includegraphics[width=0.40\textwidth]{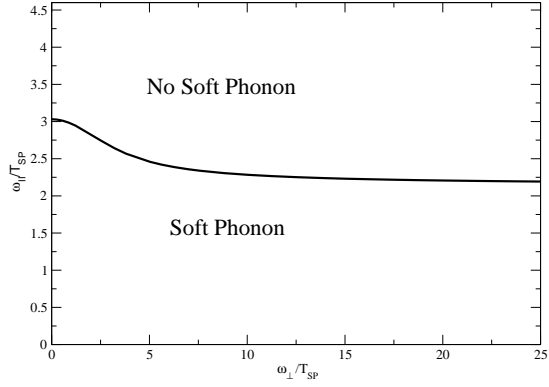}
\caption{\label{phasediag} The line separating the zone in the
$\frac{\omega_{\perp}}{T_{SP}}$,
$\frac{\omega_{\parallel}}{T_{SP}}$ plane, where the spin-Peierls
transition takes place by softening of the phonon from the one
where the Peierls active phonon does not soften completely.}
\end{figure}

The limiting value of $\omega_{\parallel} /T_{SP}$ for separation of regimes depends on
$\omega_{\perp}$. To find this transition analytically it is enough to compare concavities of both
sides of eq.\ (\ref{freres}) at $T=T_{SP}$, as seen in Fig.\ \ref{GW1}. Expanding both sides up to
second order in $\frac{\omega}{T_{SP}}$, the soft phonon regime is reached when, at $T=T_{SP}$, the
coefficient of $\omega^2$ in the r.h.s.\ is smaller than the one in the l.h.s. Under this condition
the bare phonon frequency (that obtained at hight $T$) evolves continuously to zero when the
temperature is lowered until the transition temperature. The following inequality should then be
fulfilled to have the soft phonon regime:
\begin{eqnarray}
\frac{\omega_{\parallel}}{T_{SP}} <
2\pi \sqrt{\frac{\chi_0}{\chi_2}}
\sqrt{ 1+\frac{1}{2}\frac{(1+\sqrt{1+(\omega_{\perp}/\omega_{\parallel})^2})}
{1+(\omega_{\perp}/\omega_{\parallel})^2}}.
\end{eqnarray}
In the other case, the bare phonon frequency evolves to some finite value at the SP transition.
In Fig.\ \ref{phasediag} we show the phase diagram of the spin-phonon system on the
$\omega_\perp - \omega_\parallel$ plane.
Note that in a wide range of high enough $\frac{\omega_{\perp}}{T_{SP}}$ the
separatrix is given by $\frac{\omega_{\parallel}}{T_{SP}} \sim 2.2$
as in GW. Our result extends this condition for finite values of
$\frac{\omega_{\perp}}{T_{SP}}$ where realistic spin-Peierls materials
as CuGeO$_3$ live. This also explains why the RPA approach for
the spin-phonon coupling underlying the GW calculation compares
well with experiments on CuGeO$_3$.

\begin{figure}[ht]
\vspace{2mm}
\includegraphics[width=0.40\textwidth]{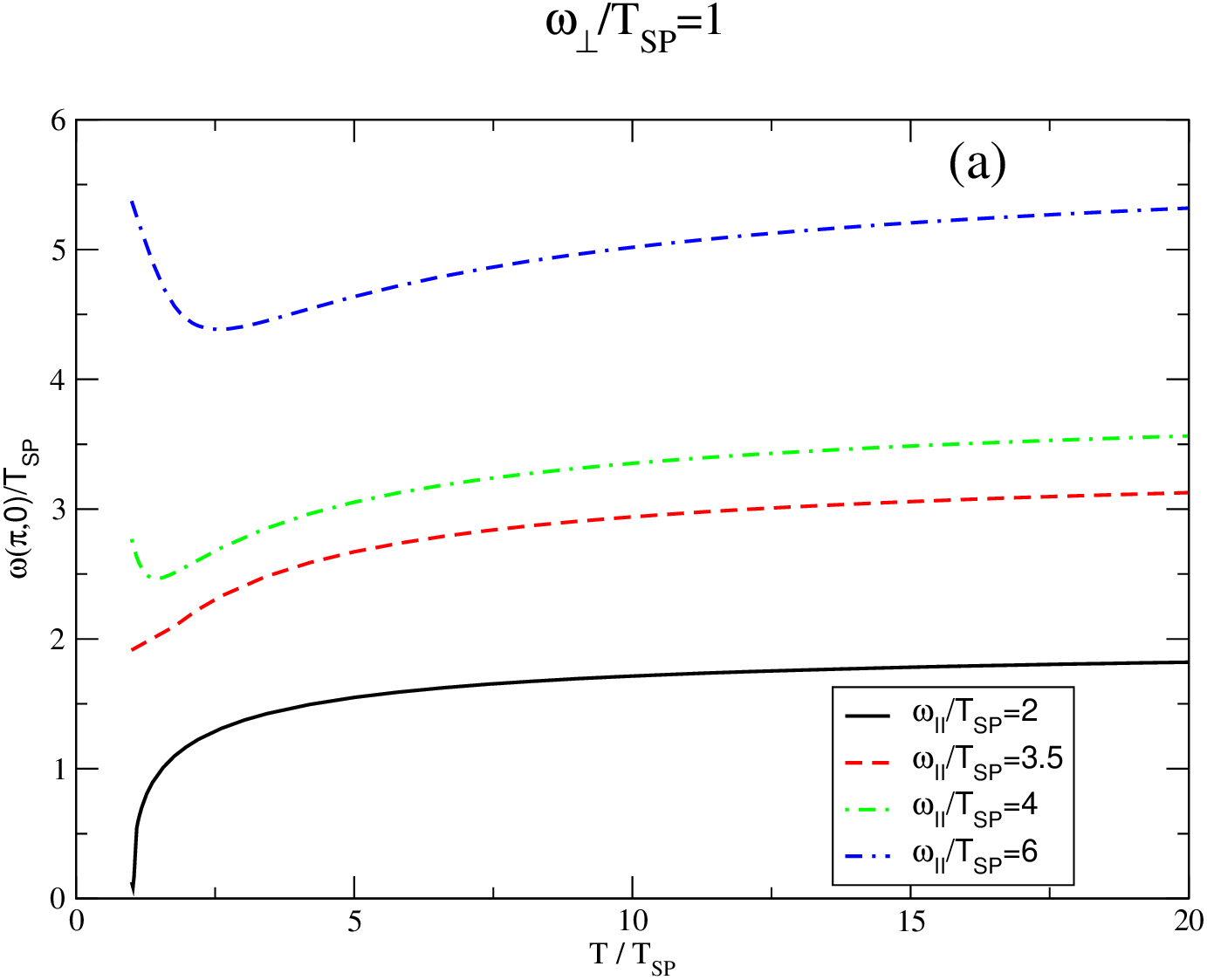}
\hspace{1mm}
\includegraphics[width=0.40\textwidth]{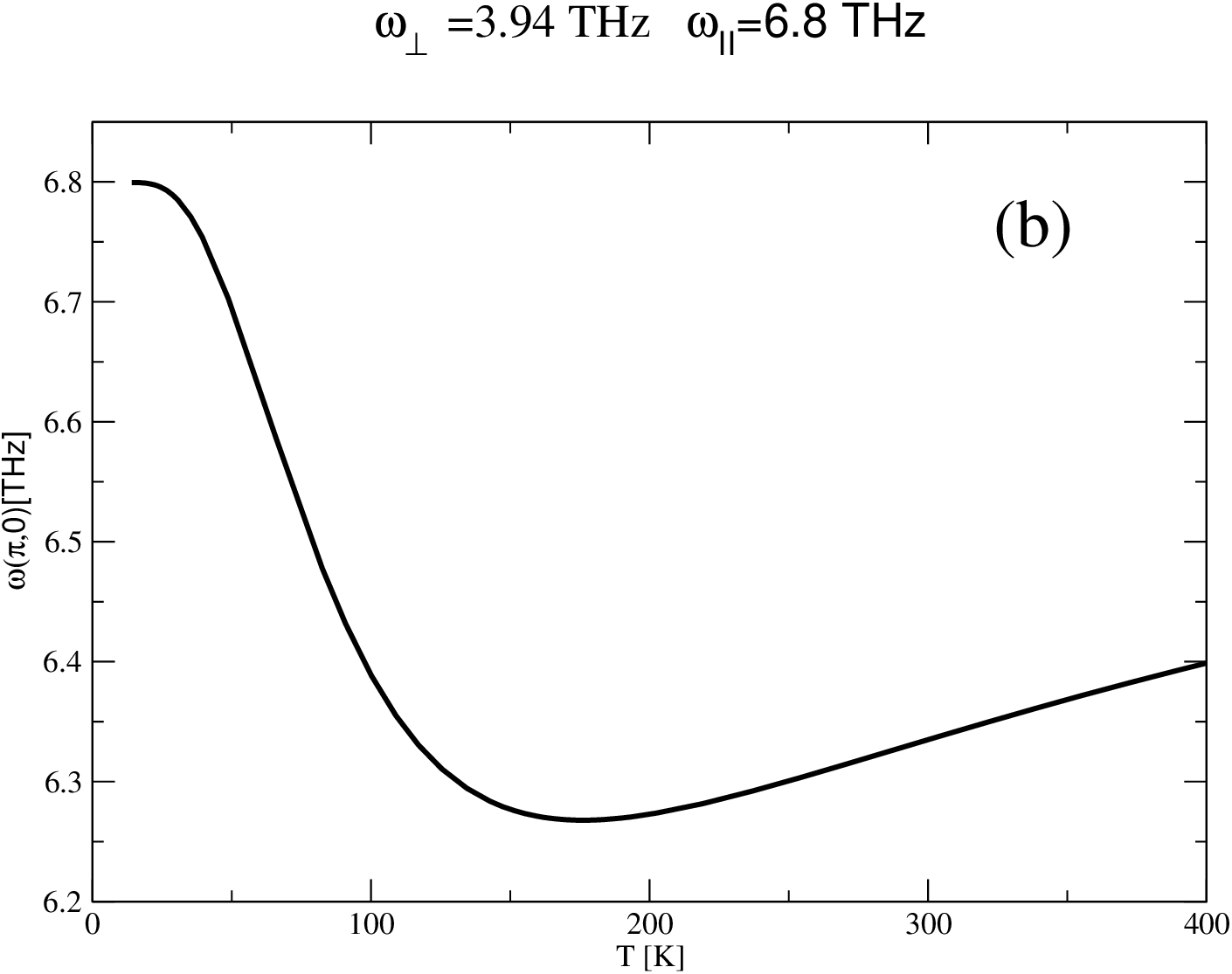}
\caption{\label{phononsvsT} (a)The temperature dependence of the
renormalized phonon frequency for different values of the bare
$\frac{\omega_{\parallel}}{T_{SP}}$, with  $\frac{\omega_{\perp}}{T_{SP}}$=1 (b) The hardening of the phonon
for the parameters describing the higher energy active SP mode in
CuGeO$_3$. The bare values used 
for $\omega_{\parallel}$ and $\omega_{\perp}$ are shown on top of the Figure.}
\end{figure}

In Fig.\ \ref{phononsvsT} we show the temperature dependence of
the renormalized phonon frequency for
$\frac{\omega_{\perp}}{T_{SP}}=1$ where the phase diagram of Fig.\
\ref{phasediag} is not completely flat. The behavior shifts from a
soft  to a non soft phonon regime at some
intermediate value of $\frac{\omega_{\parallel}}{T_{SP}}$.
In this last regime the
mechanism of the SP  transition has been associated with the
emergence of a central peak in the spectral signal measured in
neutron scatering experiments\cite{centralpeak}. For high enough
$\frac{\omega_{\parallel}}{T_{SP}}$ the phonon hardens from a
minimum value at some intermediate temperature.  At $T=T_{SP}$ the
frequency of the phonon is the bare one as in the high temperature
limit. We will use this fact to fix $\omega_{\parallel}$ for
CuGeO$_3$. Note that the $T$-dependence of the phonon is similar to
the one obtained by GW but the frequency and temperature scale are
different.

From the previous results we can
follow the $T$-dependence of the Peierls active phonons in CuGeO$_3$.
In Fig.\ \ref{phononsvsT} ({\it b}), we show our prediction for the higher energy mode
$T_2^+$, which is the most anomalous and strongly coupled to the magnetism\cite{Braden1}.
We take as a bare phonon frequency  $\omega_{\parallel}=6.8 THz$ {\em i.e.}\ its low temperature value.
For $\omega_{\perp}$ we use the relationship obtained in the previous Section.
Note that a value of
  $T_{SP}=14.1K (0.294 THz)$ has been used to give the frequencies in $Thz$ 
and the temperatures in $K$.
Our results compare well with Fig.\ 3 of [\onlinecite{Braden1}]. In
fact we predict a stronger hardening than the one experimentally
observed ($7\%$ against $4.5\%$ seen in the experiment). This
could be due to the simplified model we use, which does not take into
account the fact that two phonon modes are necessary to describe
the SP transition in CuGeO$_3$\cite{Braden1}.

Besides, we can make a prediction for
the phonon frequency behavior at higher temperature  than the one
already measured. It should increase as shown in Fig.\
\ref{phononsvsT}, approaching the bare frequency at very high $T$.
We note that at room temperature the phonon frequency does not
correspond to the bare one as expected. It is almost the minimum
of the predicted phonon frequency. The bare frequency is only
obtained at $T=T_{SP}$ and at very high temperature (probably not
experimentally accessible).

To summarize the results of the present Section, we have
generalized the MF approach of the previous Section to the
dynamical correlation functions in the high temperature phase. Our
approach relies on a RPA on the effective interchain coupling
generated by the spin-phonon interaction. In contrast to the previous
theoretical calculations, the present one does not presuppose the
existence of a soft phonon mode in the non magnetic regime. Therefore, it can been
applied to CuGeO$_3$. Moreover, the qualitative behavior of the
Peierls active phonon is similar to the one previously obtained.
There is a regime corresponding to low frequency phonon where the
structural phase transition takes place by progressive softening of
a phonon from the high temperature phase. For higher bare phonon
frequency no soft phonon is observed and  it can even harden from
the room temperature value to the transition temperature one. This is
precisely the behavior seen in neutron scattering measurements
of the phonon spectra in CuGeO$_3$.

\section{Conclusions}

We have developed a route for the spin-Peierls transition which
goes beyond the usual adiabatic treatment of the phonon field. We
emphasized the essential character that plays the dispersion of the
phonon in the transversal direction to the magnetic chains.
Moreover, the results of the present paper show that, when the
adiabatic hypothesis for the phonon coordinates is relaxed, the
one-chain model does not represent a good starting point to describe
a real system. Furthermore the effective interchain interaction
generated by the phonons is an essential ingredient for the
opening of the gap and the existence of a finite temperature phase
transition. The precise determination of the SP transition
temperature from the microscopic parameters strongly depends on the
width of the transversal phonon dispersion.

Our approach for the dynamics of the phonon mode in the high temperature
phase justifies and generalizes previous calculations of the phonon
spectra based on the Cross and Fisher early work. Applied to CuGeO$_3$,
we show that the phonon does not soften until the phase transition.
Moreover the overall temperature dependence of the phonon
frequency is consistent with experimental determinations.

As the elastic interaction between the chains play a central rol
in non-adiabatic spin-Peierls systems, we anticipate that a change
in the relative positions of the magnetic ions in different chains
will change the phase diagram and the mechanism of the phase
transition. In this sense it is highly interesting to study the
recently discovered  quasi-one-dimensional magneto-elastic system TiOCl
with an intermediate incommensurate phase where the Ti ions are not
aligned in the direction perpendicular to the magnetic
chain\cite{TiClO}. We will study in a forthcoming work this system
within the formalism developed in the present paper.

\acknowledgments We thank F.\ Essler and E.\ Orignac for helpful discussions.
This work was partially supported by ECOS-Sud Argentina-France
collaboration (Grant A04E03), PICS CNRS-CONICET (Grant 18294),
PICT ANCYPT (Grants 20350 and 03-12409), and PIP CONICET (Grants 5037 and 5306).

\end{document}